\documentclass[12pt,preprint]{aastex}
\newcommand{\beq}{\begin{equation}}
\newcommand{\beqa}{\begin{eqnarray}}
\newcommand{\eeq}{\end{equation}}
\newcommand{\eeqa}{\end{eqnarray}}
\newcommand{\simg}{\gtrsim}
\newcommand{\siml}{\lesssim}
\shorttitle{Lognormal Distributions in Gamma-Ray Bursts}
\shortauthors{Ioka \& Nakamura}
\begin{document}
\title{
Possible Origin of Lognormal Distributions in Gamma-Ray Bursts
}
\author{
Kunihito Ioka$^{1}$
and Takashi Nakamura$^{2}$
}
\affil{$^{1}$ Department of Earth and Space Science,
Osaka University, Toyonaka 560-0043, Japan}
\affil{
$^{2}$Yukawa Institute for Theoretical Physics, Kyoto University, 
Kyoto 606-8502, Japan}
%\slugcomment{Feb~~2~~2002; Mar~~20~~2002, OU-TAP 171,YITP-}
\email{ioka@vega.ess.sci.osaka-u.ac.jp,
takashi@yukawa.kyoto-u.ac.jp}

\begin{abstract}
We show that if the intrinsic break energy of Gamma-Ray Bursts
(GRBs) is determined
by the product of more than three random variables
the observed break energy distribution becomes almost lognormal including
the redshift effect because of the central limit theorem.
The excess from the lognormal distribution at the low break energy
is possibly due to the high redshift GRBs.
The same argument may also apply to the pulse duration, 
the interval between pulses and so on.
\end{abstract}


\section{INTRODUCTION}\label{sec:intro}
Among the statistical properties of the observed quantities,
lognormal distributions are frequently seen in Gamma-Ray Bursts (GRBs).
The lognormal distribution may be defined as the distribution
of a random variable $x$ whose logarithm is normally distributed,
\beqa
f(x)dx = \left\{
\begin{array}{ll}
{\frac{1}{\sqrt{2\pi} \sigma}} 
\exp{\left[-\frac{(\log x- \mu)^2}{2\sigma^2}\right]} d \log x,& 
\quad {\rm if} \quad x>0,\\
0,& \quad {\rm if} \quad x \le 0,
\end{array}\right.
\eeqa
where $f(x)$ is the probability density function for $x$, and
$\mu$ and $\sigma^2$ are the sample mean and the variance of $\log x$
(e.g., Crow \& Shimizu 1988).
This distribution is unimodal and positively skew.
McBreen, Hurley, Long \& Metcalfe (1994)
pointed out that the total duration of the long and short bursts
and the time interval between pulses are consistent with the
lognormal distributions.
Li \& Fenimore (1996) showed that the pulse fluence
and the pulse interval distributions within each burst are
consistent with lognormal distributions.
Nakar \& Piran (2001) found that the pulse duration also
have a lognormal distribution.
The break energy distribution is also lognormal
(Preece et al. 2000; see below). 

Lloyd, Fryer \& Ramirez-Ruiz (2001) suggested that
$\sim$ 10\% of GRBs might have the redshift larger than 6
so that the redshift distribution might be wide.
Therefore it is quite strange that the observed break 
energy distribution and the duration distribution are lognormal
since it does not seem that the observed lognormal distribution 
reflects the redshift distribution of the GRBs.
Even if the break energy distribution is lognormal at the source,
the observed break energy should be smaller than 
the intrinsic one by a factor of $(1+z)$ while
the observed duration should be longer than the intrinsic one
by a factor of $(1+z)$.
These factors change by order of unity between $z=0$ and $z\sim 6$.

In this Letter, we will consider a possible origin
of the observed lognormal distributions in GRBs
from the viewpoint of the central limit theorem.\footnote{The lognormal 
distribution of the time interval between pulses and the pulse fluence might
be reproduced by a fine tuning of model parameters
(Spada, Panaitescu \& M${\acute {\rm e}}$sz${\acute {\rm a}}$ros 2000).}

\section{BREAK ENERGY DISTRIBUTIONS}
\begin{figure}
\plotone{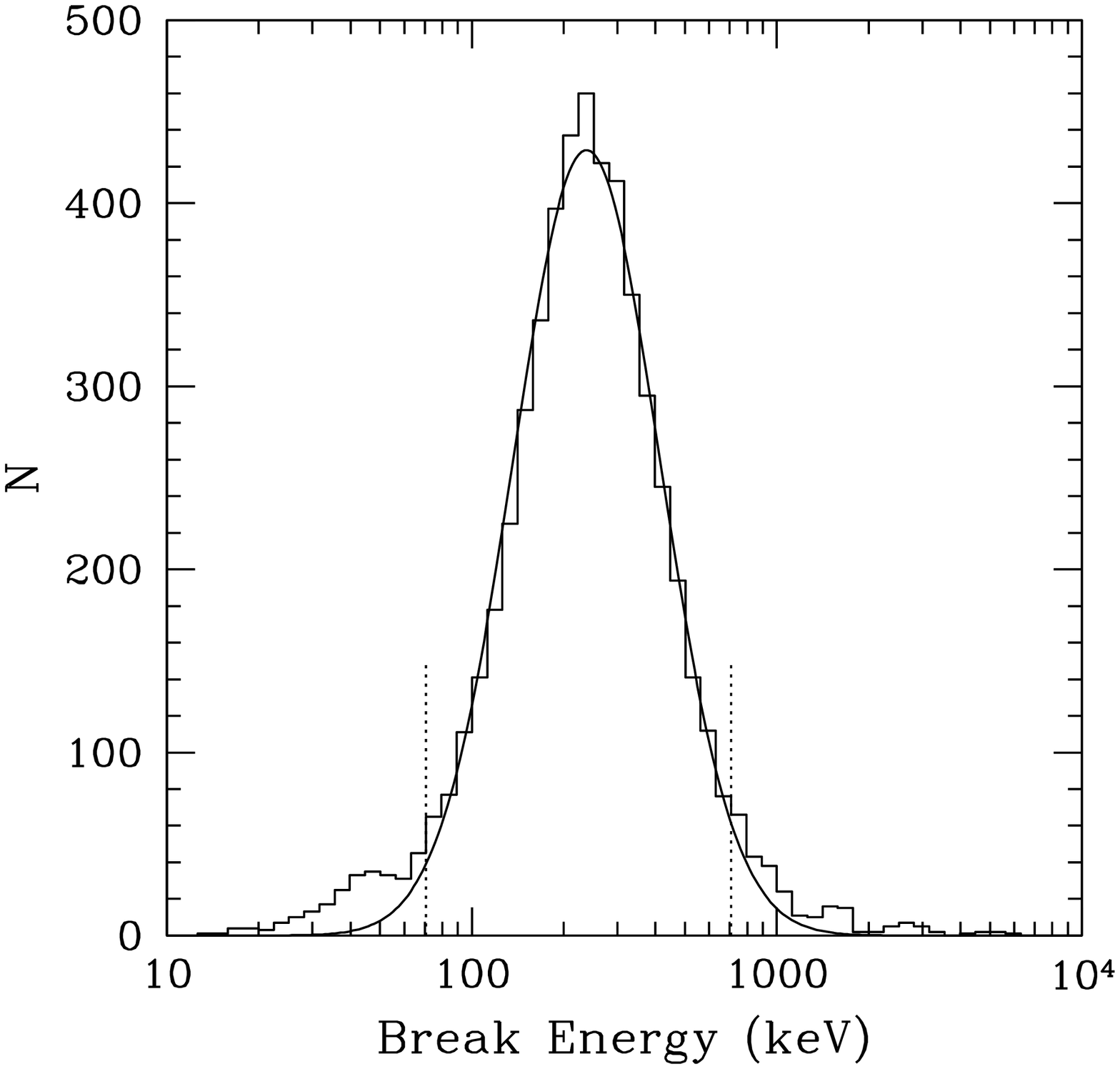}
\caption{The break energy distribution 
taken from the electronic edition of Preece et al. (2000) is shown.
The lognormal fitting 
of the data between $70.8$ keV and $708$ keV
is shown by a solid line.
The fitting range between $70.8$ keV and $708$ keV
is shown by dotted lines.
The $\chi^2$ test gives the probability of 0.497 
(the reduced $\chi^2$ is 0.963 with 17 degrees of freedom) 
that the data was taken from the lognormal distribution,
with the mean $\mu=2.38 \pm 0.004$ ($E_{b}\simeq 238$ keV)
and the standard deviation
$\sigma=0.240 \pm 0.004$ ($1\sigma$ width is between $137$ keV 
and $413$ keV).} \label{fig:be}
\end{figure}

Figure \ref{fig:be} shows the histogram of the break energy $E_b$
taken from the electronic edition of Preece et al. (2000).\footnote{
It is not known whether the paucity of the soft and hard bursts
is real or not, because harder bursts have fewer photons
(Cohen, Piran \& Narayan 1998; Lloyd \& Petrosian 1999;
but see Brainerd et al. 1999)
and there may exist relatively many soft bursts
with low luminosities, so called X-ray rich GRBs
(or X-ray flashes or Fast X-ray transients)
(Strohmayer et al. 1998; Heise et al. 2001; Kippen et al. 2001).
Here we assume that the selection effect is small.}
The $\chi^2$ test of all data gives the probability
of $1.4 \times 10^{-185}$ 
(the reduced $\chi^2$ is 16.5 with 66 degrees of freedom)
that the data was taken from the lognormal distribution.
Therefore the null hypothesis that the break energy distribution
is lognormal fails.
However, if we exclude the data in the high and low energy ends,
the fit becomes good as shown in Figure \ref{fig:be}.
The $\chi^2$ test of the data between $70.8$ keV and $708$ keV
gives the probability of 0.497 
(the reduced $\chi^2$ is 0.963 with 17 degrees of freedom) 
that the data was taken from the lognormal distribution,
with $\mu=2.38 \pm 0.004$ ($E_{b}\simeq 238$ keV)
and $\sigma=0.240 \pm 0.004$ ($1\sigma$ width is between $137$ keV 
and $413$ keV).
The improvement of the lognormal fit
to the break energy distribution excluding the high and low energy ends
may suggest that the soft and hard bursts originate 
from a different class of GRBs or emission mechanisms.
Anyway, hereafter we will assume that the observed break energy 
distribution is lognormal.

Now let us assume that the intrinsic break energy distribution
is lognormal. 
Next we numerically calculate 
the observed break energy distribution
assuming that the redshift distribution has the form,
\beqa
f(z)dz=\left\{
\begin{array}{ll}
A (1+z)^{a-1} (1+z_0)^{b-1}dz,& \quad {\rm if}\quad 0<z<z_0,\\
A (1+z_0)^{a-1} (1+z)^{b-1}dz,& \quad {\rm if}\quad z_0<z,
\end{array}\right.
\label{eq:fz}
\eeqa
where $b<0$.
This redshift distribution
%\footnote{
%The normalization in equation (\ref{eq:fz}) can be calculated as
%\beqa
%A&=&[(1+z_0)^{a+b-1}(a^{-1}-b^{-1})-(1+z_0)^{b-1}a^{-1}]^{-1},
%\eeqa
%and the mean and the variance of $\log (1+z)$ are
%\beqa
%\mu_z&=&\frac{(1+z_0)^{a}(a^{-1}-b^{-1})[\ln (1+z_0)-a^{-1}-b^{-1}]+a^{-2}}
%{(\ln 10)[(1+z_0)^{a} (a^{-1}-b^{-1})-a^{-1}]},
%\\
%\sigma_z^2&=&\bigl[(1+z_0)^{2a}(a^{-1}-b^{-1})^2 (a^{-2}+b^{-2})
%+a^{-4}
%\nonumber\\
%&-&(1+z_0)^{a} a^{-1}(a^{-1}-b^{-1}) 
%\left\{[\ln (1+z_0)]^2-2 b^{-1} \ln (1+z_0)+2(a^{-2}+b^{-2})\right\}
%\bigr]
%\nonumber\\
%&\times&{(\ln 10)^{-2} [(1+z_0)^{a} (a^{-1}-b^{-1})-a^{-1}]^{-2}}.
%\eeqa
%A fraction of the GRBs at $z>z_1(>z_0)$ is given by
%$A b^{-1} (1+z_0)^{a-1} (1+z_1)^{b}$,
%which is $9.4$ \% for $(z_0,z_1,a,b)=(3,6,3,-3)$
%for example.
%}
rises as $\propto (1+z)^{a-1}$ 
to a redshift of $z_0$
and then declines as $\propto (1+z)^{b-1}$,
and it is similar to that in
Figure 8 of Lloyd, Fryer \& Ramirez-Ruiz (2001).
To mimic Figure \ref{fig:be}, 
we generate 155 bursts with each burst having 35 spectra
for each realization.
We take the mean and the variance of the intrinsic break energy distribution
%as $\mu=2.38+\xi \mu_z$ and $\sigma^2=(0.240)^2-(\zeta \sigma_z)^2$
%with $\xi$ and $\zeta$ being constants,
so that the mean and the variance of the observed break energy 
are close to that in Figure \ref{fig:be}.
%, where $\mu_z$ and $\sigma_z^2$
%are the mean and the variance of $\log (1+z)$, respectively.\footnote{
%Since we truncate the data in the high and the low energy ends
%as in Figure \ref{fig:be},
%the mean and the variance of the observed break energy distribution
%do not coincide with $\mu=2.38$ and $\sigma^2=(0.240)^2$, respectively,
%if we take the mean and the variance of the intrinsic break energy 
%distribution
%as $\mu=2.38+\mu_z$ and $\sigma^2=(0.240)^2-\sigma_z^2$, respectively.
%We estimate the factors, $\xi$ and $\zeta$,
%in front of $\mu_z$ and $\sigma_z$ through trial and error,
%so that the mean and the variance of the observed break energy 
%are close to that in Figure \ref{fig:be}.
%We use $(\xi,\zeta)=(0.9,0.8)$ for $z_0=0$,
%$(\xi,\zeta)=(1.0,0.9)$ for $(z_0,a)=(3,2)$,
%$(\xi,\zeta)=(1.0,0.8)$ for $(z_0,a)=(3,3)$
%and $(\xi,\zeta)=(1.0,0.8)$ for $(z_0,a)=(3,5)$.
%But the main conclusion does not depend on the precise 
%values of these factors.}
For each realization, we make a $\chi^2$ fitting
to obtain the probability 
that the data is taken from the lognormal distribution.
As in Figure \ref{fig:be}, we use the data between 70.8 keV and 708 keV 
for the $\chi^2$ test.
%Figure \ref{fig:ap} 
%shows the average probability for $10^4$ experimental realizations
%as a function of the power law index $-b$.
%Some irregular features are the statistical fluctuation.
%The average probabilities fall off at $-b \siml 2.5$
%because the variances of the redshift distribution $\sigma_z^2$
%approach the observed one $(0.240)^2$.
%The variances of the redshift distribution $\sigma_z^2$
%at $-b=4$ are $(0.109)^2$, $(0.191)^2$, $(0.168)^2$ and $(0.138)^2$
%for $z_0=0$, $(z_0,a)=(3,2)$, $(z_0,a)=(3,3)$ and $(z_0,a)=(3,5)$,
%respectively.
%
%Therefore this simulation shows that
From this simulation, we have found that
the average probability can reach $\sim 0.5$
even when the variance of the redshift distribution 
is comparable to the observed one $\sigma_z^2 \sim (0.240)^2$,
although it is slightly stronger condition to preserve
the lognormal form than $\sigma_z^2 < (0.240)^2$.
At first glance it is strange that the simulations do not reflect the redshift
distribution contrary to the argument in Section \ref{sec:intro}.
In the next section we will show that it is not strange but natural
due to the central limit theorem.

\begin{figure}
\plotone{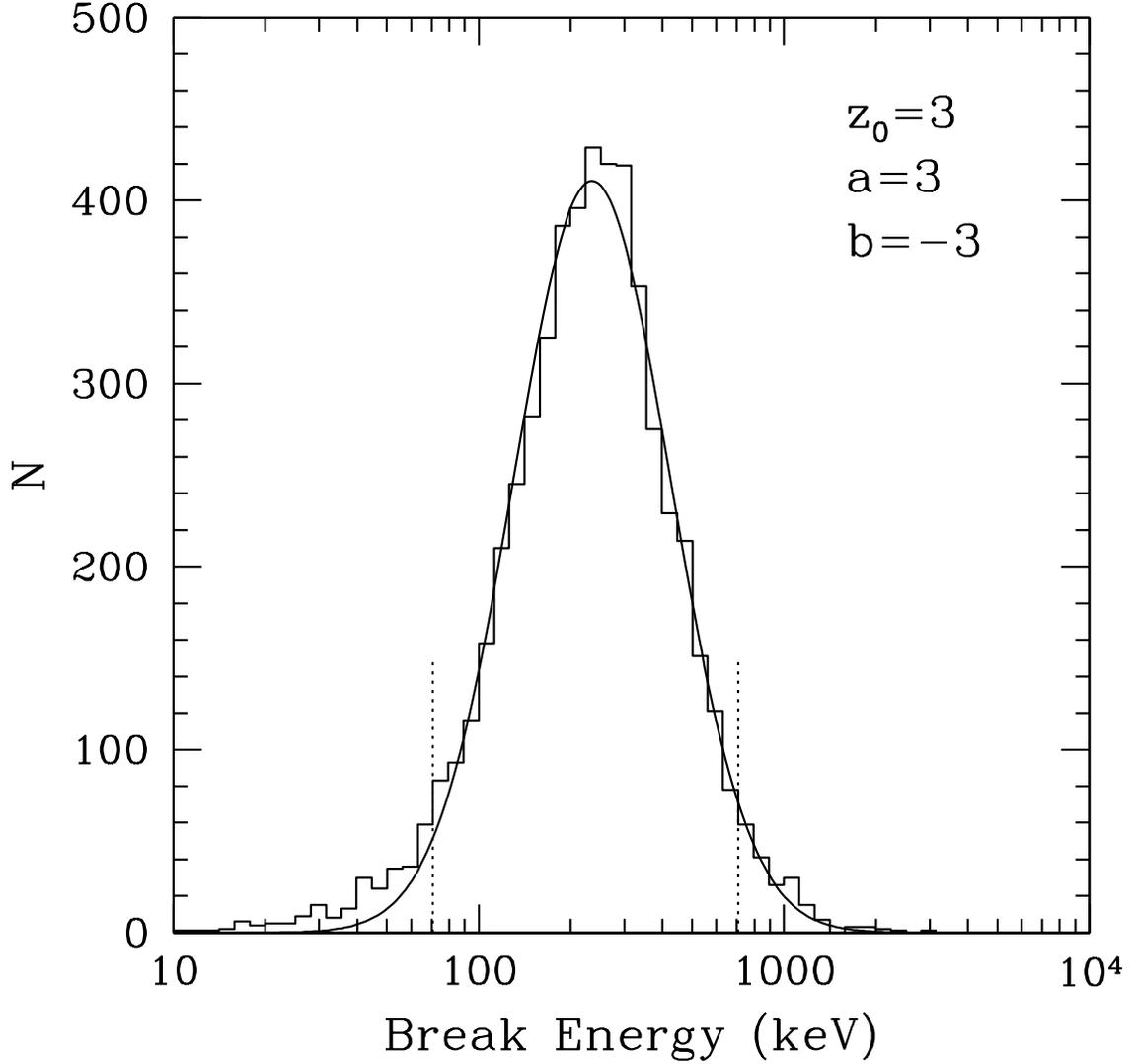}
\caption{One experimental realization of the observed break energy distribution
for $(z_0,a,b)=(3,3,-3)$ in equation (\ref{eq:fz}) is shown.
The lognormal fitting 
of the data between $70.8$ keV and $708$ keV
is shown by a solid line.
The fitting range between $70.8$ keV and $708$ keV
is shown by dotted lines.
The $\chi^2$ test gives the probability of $0.128$
(the reduced $\chi^2$ is $1.39$ with $17$ degrees of freedom) 
that the data was taken from the lognormal distribution,
with $\mu=2.37 \pm 0.004$ ($E_{b}\simeq 235$ keV)
and $\sigma=0.256 \pm 0.004$ ($1\sigma$ width is between $130$ keV 
and $423$ keV).
There is an excess of soft bursts
relative to the lognormal fit as in Figure \ref{fig:be}. \label{fig:besim}}
\end{figure}

Figure \ref{fig:besim} shows the histogram of the observed break energy
for one experimental realization with $(z_0,a,b)=(3,3,-3)$.
The $\chi^2$ test of the data between $70.8$ keV and $708$ keV
gives the probability of $0.128$
(the reduced $\chi^2$ is $1.39$ with $17$ degrees of freedom) 
that the data was taken from the lognormal distribution,
with $\mu=2.37 \pm 0.004$ ($E_{b}\simeq 235$ keV)
and $\sigma=0.256 \pm 0.004$ ($1\sigma$ width is between $130$ keV 
and $423$ keV).
It is interesting to note the excess of soft bursts
relative to the lognormal fit as in Figure \ref{fig:be}.
The average redshifts of these soft bursts are relatively high.

\section{LOGNORMAL DISTRIBUTIONS}
The standard model of the GRB emission
is the optically thin synchrotron shock model (e.g., Piran 1999).
A similar discussion in the following
will be applied to the inverse Compton model.
Let us consider a slow (rapid) shell
with a Lorentz factor $\gamma_s$ ($\gamma_r$),
a mass $m_s$ ($m_r$) and a width $l_s$ ($l_r$).
When a separation between two shells is $L$,
the collision takes place at a radius $R_s \simeq 2 L \gamma_s^2$.
At the collision,
the forward and the reverse shock are formed.
Here we consider the reverse shock propagating into the rapid shell.
The discussion for the forward shock is similar.
We assume that a fraction of electrons $\zeta_e$ is accelerated
in the shock to a power law distribution of Lorentz factor $\gamma_e$,
$N(\gamma_e) d\gamma_e \propto \gamma_e^{-p} d\gamma_e$
for $\gamma_e \ge \gamma_{min} \equiv
[(p-2)/(p-1)](\epsilon_e u'/\zeta_e n' m_e c^2)$,
where $n'$ and $u'$ are the number density and the internal energy
density in the local frame, respectively, $p \simg 2$, and
we assume that a fraction $\epsilon_e$ of the internal energy
goes into the electrons.
We also assume that a fraction $\epsilon_B$ of the internal energy
goes into the magnetic field, $B^2=8\pi \epsilon_B u'$.
The local frame quantities, $u'$ and $n'$, 
can be calculated using the shock jump conditions
(Blandford \& Mckee 1976; Sari \& Piran 1995).
We assume that the unshocked shells are cold
and the shocked shells are extremely hot.
If the Lorentz factor of the shocked region is $\gamma$,
the relative Lorentz factor of the unshocked and the shocked region
is given by 
$\gamma_{rel} \simeq (\gamma_r/\gamma+\gamma/\gamma_r)/2
\simeq \gamma_r/2\gamma$,
so that $u'=(\gamma_{rel}-1) n' m_p c^2 \simeq \gamma_{rel} n' m_p c^2$.
The number densities of the unshocked and the shocked region
are given by $n_r'=m_r/4\pi m_p R_s^2 l_r \gamma_r
\simeq m_r/16 \pi m_p L^2 \gamma_s^2 l_r \gamma_r$
and $n'=(4\gamma_{rel}+3) n_r' \simeq 4\gamma_{rel} n_r'$, respectively.
Thus, the characteristic synchrotron energy is given by
\beqa
E_b={{\hbar q_e B \gamma \gamma_{min}^2}\over{m_e c (1+z)}}\simeq
260 \left[{{p-2}\over{p-1}}\right]^2 
\epsilon_e^2 \epsilon_B^{1/2} \zeta_e^{-2} m_{r,28}^{1/2} 
L_{10}^{-1} l_{r,9}^{-1/2} \gamma_{s,2}^{-2} \gamma_{r,2}^{1/2}
\gamma_{rel}^2 (1+z)^{-1} \ {\rm keV},
\label{eq:benergy}
\eeqa
where we assume that the source is at a redshift $z$.
Note that the relative Lorentz factor of the unshocked and shocked region
$\gamma_{rel}$ depends on the relative Lorentz factor
of the rapid and slow shell $\gamma_{rs} 
\simeq (\gamma_r/\gamma_s+\gamma_s/\gamma_r)/2
\simeq \gamma_r/2 \gamma_s$ and 
the ratio between the number densities in these shells
$f \equiv  n_s'/n_r'=m_s l_r \gamma_r/m_r l_s \gamma_s$
(Sari \& Piran 1995).
For the ultrarelativistic shock case $\gamma_{rs}^2 \gg f$,
$\gamma_{rel}=\gamma_{rs}^{1/2} f^{1/4}/\sqrt{2}
=(m_s l_r \gamma_r^3/m_r l_s \gamma_s^3)^{1/4}/2$.

Equation (\ref{eq:benergy}) shows that
the break energy is written in the form of a product of many variables.
For such a variable made from the product of many variables,
the lognormal distribution may have a very simple origin, that is,
the central limit theorem (Crow \& Shimizu 1988;
Montroll \& Shlesinger 1982).
Let a variable $q$ be written in the form of a product of variables,
\beqa
q=x_1 x_2 \cdots x_n.
\eeqa
Then,
\beqa
\log q=\log x_1 + \log x_2 + \cdots + \log x_n.
\eeqa
When the individual distributions of $\log x_i$
satisfy certain weak conditions that include the existence of second moments,
the central limit theorem is applicable to the variable $\log q$, so that
the distribution function of $\log q$ tends to the normal distribution
as $n$ tends to infinity.

\begin{figure}
\plotone{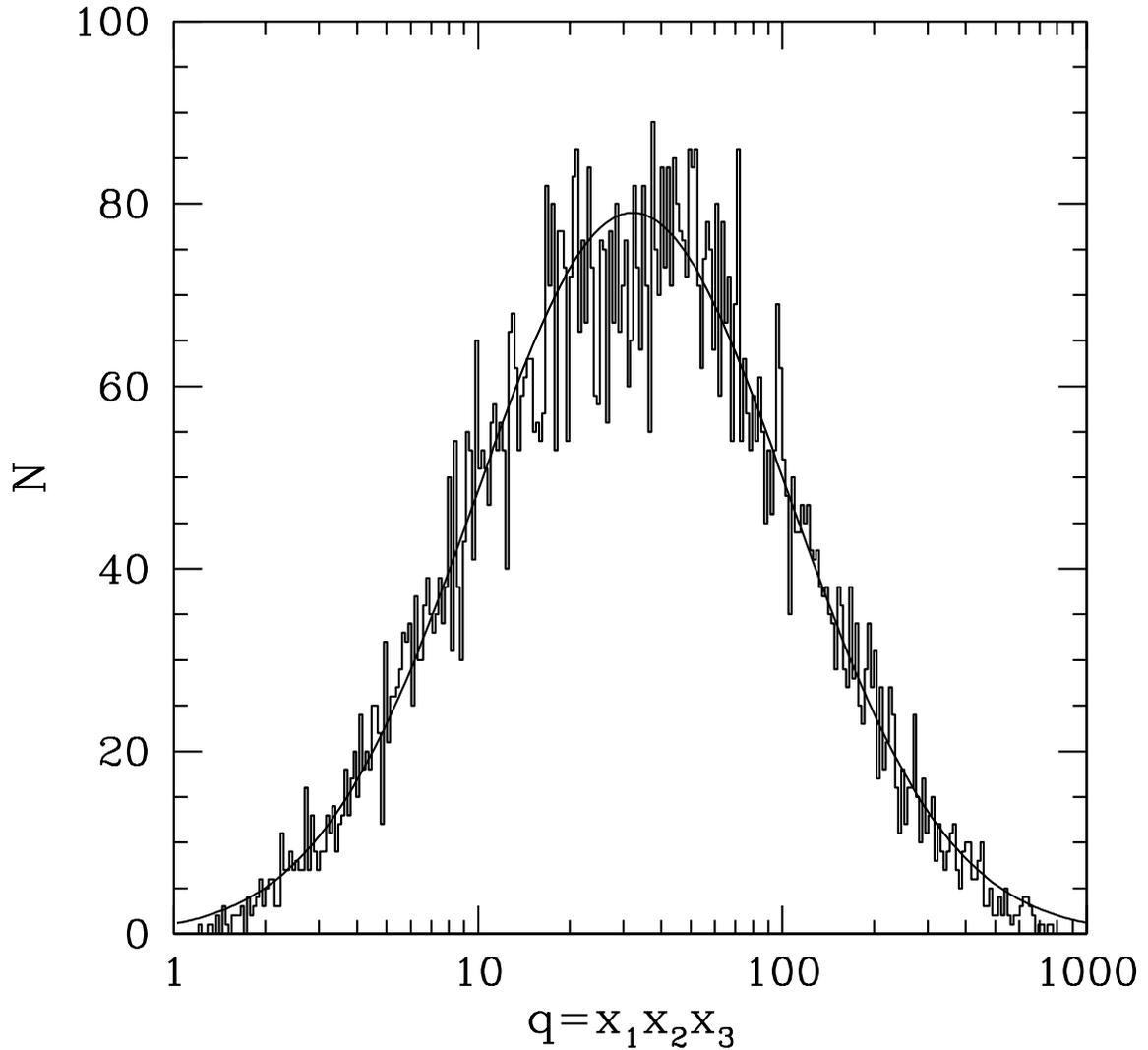}
\caption{The distribution of the product
of three random variables $q=x_1 x_2 x_3$,
each distributing uniformly between 0 and 1
in the logarithmic space (that is $0<\log x_i<1$),
for $10^4$ experimental realizations is shown.
The lognormal fitting 
of the data is shown by a solid line.
The $\chi^2$ test gives the probability of $0.483$ 
(the reduced $\chi^2$ is 1.00 with 278 degrees of freedom) 
that the distribution of $q$ is taken from the lognormal distribution. 
\label{fig:three}}
\end{figure}

As an example, we numerically generated random variables $x_i$ $(i=1,2,\cdots)$
whose logarithms are uniformly distributed between 0 and 1.
Figure \ref{fig:three} shows the histogram of the product
of these three variables, $q=x_1 x_2 x_3$, 
for $10^4$ experimental realizations.
The distribution of $q$ 
agrees with the lognormal distribution quite well.
It is surprising that the $\chi^2$ test gives the probability of $0.483$ 
(the reduced $\chi^2$ is 1.00 with 278 degrees of freedom) 
that the distribution of $q$ is taken from the lognormal distribution.
This example shows that the lognormal distributions
may be achieved by a relatively small number of variables
(Yonetoku \& Murakami 2001).
Note that when the number of the variables is two, i.e., $q=x_1 x_2$,
the probability 
that the distribution is taken from the lognormal distribution
was only $1.64 \times 10^{-5}$,
so that a product of only one more variable may make a distribution lognormal.

Therefore the lognormal distribution of the break energy
may be a natural result from the central limit theorem.
We may say, ``Astrophysically, not $n=\infty$ but $n=3$
gives the lognormal distribution !!''.
The effect of the redshift is just to add one variable in 
equation (\ref{eq:benergy})
if the redshift in the observed data is randomly chosen.

\section{PULSE FLUENCE/DURATION/INTERVAL DISTRIBUTIONS}
Let us consider the lognormal distributions
in other quantities related to GRBs.
When the rapid shell catches up the slow one in the internal shock,
using the conservation of the energy and the momentum,
the Lorentz factor of the merged shell $\gamma_m$
and the internal energy $E_{int}$ produced by the collision
are given by
$\gamma_m \simeq [(m_r \gamma_r+m_s \gamma_s)/
(m_r/\gamma_r+m_s/\gamma_s)]^{1/2}$
and $E_{int}=m_r \gamma_r + m_s \gamma_s -(m_r+m_s) \gamma_m$,
respectively (e.g., Piran 1999).
If we assume that a fraction $\epsilon_e$ of the internal energy
goes into the electrons
and a fraction $\epsilon_w$ of the energy radiated by the electrons
is within the gamma-ray band,
the observed energy is given by
\beqa
E_{obs}=\epsilon_w \epsilon_e E_{int} (1+z)^{-1}\sim
\epsilon_w \epsilon_e m_r 
\gamma_r (1+z)^{-1}.
\label{eq:eobs}
\eeqa
Equation (\ref{eq:eobs}) shows that the observed energy,
which is proportional to the pulse fluence,
is written in the form of a product of five variables,\footnote{
Although the fraction $\epsilon_e$ of the internal energy
that goes into electrons may be a fundamental constant,
there is still a large dispersion of order unity in $\epsilon_e$
deduced from the the afterglow observations (Panaitescu \& Kumar 2001).}
$\epsilon_w$, $\epsilon_e$, $m_r$,
$[\gamma_r-\gamma_m+(m_s/m_r) (\gamma_s-\gamma_m)]\sim \gamma_r$
and $(1+z)^{-1}$.
Therefore the lognormal distribution of the pulse fluence
may be a natural result from the central limit theorem.

The pulse duration is determined by three time scales:
the hydrodynamic time scale, the cooling time scale,
and the angular spreading time scale 
(Kobayashi, Piran \& Sari 1997; Katz 1997; Fenimore, Madras \& Sergei 1996).
The cooling time scale is usually much shorter than 
the other two time scales in the internal shocks
(Sari, Narayan \& Piran 1996).
The hydrodynamic time scale $\sim l/c$ and the angular spreading
time scale determine the rise and the decay time of the pulse, respectively.
Since most observed pulses rise more quickly than they decay
(Norris et al. 1996), we assume that
the pulse duration is mainly determined by the angular spreading time
$\simeq R_s/2 c \gamma_m^2$.
Then, the pulse duration $\delta t$ is given by
\beq
\delta t \simeq (L/c) (\gamma_s^2/\gamma_m^{2}) (1+z).
\label{eq:pdur}
\eeq
On the other hand,
the interval between pulses $\Delta t$
is determined by the separation between shells,
\beqa
\Delta t \simeq (L/c)(1+z),
\label{eq:dt}
\eeqa
since all shells are moving towards us with almost the speed of light
(Kobayashi, Piran \& Sari 1997; Nakar \& Piran 2001).

Equation (\ref{eq:dt}) shows that
the pulse interval $\Delta t$ reflects
the separation between shells $L$,
while equation (\ref{eq:pdur}) shows that
the pulse duration $\delta t$ is multiplied by one more factor
$(\gamma_s/\gamma_m)^2$ other than $(L/c)(1+z)$.
Therefore, if we consider that the distribution of a product of variables
tends to the lognormal distribution
as the number of the multiplied variables increases,
the distribution of the pulse duration $\delta t$
may be closer to the lognormal distribution than that of the pulse interval
$\Delta t$.
In fact, Nakar \& Piran (2001) argued that
the pulse duration $\delta t$ has the lognormal distribution
while the pulse interval $\Delta t$ does not,
as noticed by Li \& Fenimore (1996).
The pulse interval has an excess of long intervals
relative to the lognormal distribution.
This may suggest the existence of a different distribution, i.e.,
quiescent times (long periods with no activity)
(Nakar \& Piran 2001; Ramirez-Ruiz \& Merloni 2001).\footnote{
The correlation between the pulse interval
$\delta t$ and the pulse duration $\Delta t$
is broken when the separation 
between shells is too large 
to collide each other before the external shock.
However, almost all quiescent times in Nakar \& Piran (2001)
are smaller than the limit
$\delta t \siml 100\ {\rm sec}\ E_{52}^{1/3} n_1^{-1/3} \gamma_{100}^{-8/3}$
so that shells may collide.}
But the central limit theorem may be also responsible
for the lognormal distribution of the pulse duration
$\delta t$.

\section{DISCUSSIONS}
We considered the possible origin of the lognormal distributions
in the break energy, the pulse fluence and the pulse duration
as a result of the central limit theorem.
Astrophysically the lognormal distribution may be achieved by 
a product of only a few variables.
The effect of the redshift is just to add one variable to the product
so that the redshift distribution is hidden.

We have no idea about the origin of the lognormal distributions in
the pulse interval\footnote{
Note that the distribution of the pulse interval
is lognormal if we exclude the quiescent times (Nakar \& Piran 2001).}
$\Delta t$ and the total duration $\Delta T$.
However the viewing angle may be one factor to be multiplied to 
the pulse interval $\Delta t$.
Recently we suggested that the luminosity-lag relation
could be explained by the variation in the viewing angle $\theta_v$
from the axis of the jet (Ioka \& Nakamura 2001; Nakamura 2000).
The duration of the pulse from the jet also depends on the viewing angle,
and according to Figure 2 of Ioka \& Nakamura (2001) we have
\beqa
\Delta t \propto (L/c) (1+z) (1+\gamma^2 \theta_v^2),
\eeqa
when $\theta_v \sim \Delta \theta$
where $\gamma$ is the Lorentz factor of the jet and
$\Delta \theta$ is the opening half-angle of the jet.
The multiplied factor $(1+\gamma^2 \theta_v^2)$
may be responsible for the lognormal distribution of the pulse interval
$\Delta t$.
Note that the total duration $\Delta T$ is equal to 
the lifetime of the central engine
and thus does not depend on $\theta_v$.

%We have not taken the redshift evolution of the quantities into account.
%If the redshifts of GRBs were measured,
%we could argue the correlation of observed quantities with the redshift.
%We may use some distance indicators, such as
%luminosity-lag relation (Norris, Marani \& Bonnell 2000;
%Schaefer, Deng \& Band 2001),
%and luminosity-variability relation (Fenimore \& Ramirez-Ruiz 2000;
%Reichart et al. 2000).
%This is an interesting future problem.
%It is interesting to note that
%the evolution of the GRB luminosity function was
%suggested using the luminosity-variability relations
%(Lloyd, Fryer \& Ramirez-Ruiz 2001).
%The number of physical variables is actually finite,
%and hence, it is important to investigate
%the departures from the lognormal form by the future observations.

\acknowledgments
We are grateful T. Murakami, D. Yonetoku and S. Kobayashi for useful comments.
We thank the Yukawa Institute for Theoretical Physics at Kyoto University,
where this work was initiated during the YITP-W-01-06 on ``GRB2001''.
This work was supported in part by
Grant-in-Aid for Scientific Research Fellowship
of the Japanese Ministry of Education,
Science, Sports and Culture, No.00660 (KI) and by
Grant-in-Aid of Scientific Research of the Ministry of Education,
Culture, and Sports, No.11640274 (TN) and 09NP0801 (TN).

%
% Figure captions
%

\newpage 

%\newpage 
%\begin{figure}
%\plotone{f2.eps}
%\caption{The average probability 
%that the simulated data is taken from the lognormal distribution
%for $10^4$ experimental realizations
%is shown as a function of the power law index 
%of the redshift distribution $-b$ in equation (\ref{eq:fz}).
%The simulated data is generated by calculating the observed
%break energy distribution 
%assuming that the intrinsic break energy distribution is
%lognormal and the redshift distibution has the form in equation 
%(\ref{eq:fz}).
%Four cases are shown, $z_0=0$, $(z_0,a)=(3,2)$,
%$(z_0,a)=(3,3)$ and $(z_0,a)=(3,5)$. \label{fig:ap}}
%\end{figure}

\newpage 

\newpage 

\end{document}